\documentclass[12pt]{article}
\usepackage{epsf}
\def\be{\begin{equation}}
\def\ee{\end{equation}}
\def\bea{\begin{eqnarray}}
\def\eea{\end{eqnarray}}

\def\yp{\Upsilon}

\begin{document}
\begin{titlepage}
\begin{center}
{\Large \bf William I. Fine Theoretical Physics Institute \\
University of Minnesota \\}
\end{center}
\vspace{0.2in}
\begin{flushright}
FTPI-MINN-12/12 \\
UMN-TH-3039/12 \\
April 2012 \\
\end{flushright}
\vspace{0.3in}
\begin{center}
{\Large \bf Charged-to-neutral heavy meson yield ratio at the $Z_b^0$ resonances
as a probe of the $I^G(J^P)=0^-(1^+)$ channel.\\}
\vspace{0.2in}
{\bf M.B. Voloshin  \\ }
William I. Fine Theoretical Physics Institute, University of
Minnesota,\\ Minneapolis, MN 55455, USA \\
and \\
Institute of Theoretical and Experimental Physics, Moscow, 117218, Russia
\\[0.2in]

\end{center}

\vspace{0.2in}

\begin{abstract}
It is argued that the ratio of the yield of pairs of charged and neutral $B^{(*)}$ mesons in the processes $\Upsilon(5S) \to \pi^0 + (B \bar B^* + c.c.)$ and $\Upsilon(5S) \to \pi^0 + B^* \bar B^*$ is very sensitive near the corresponding heavy meson threshold to the strong interaction between the mesons in the $I^G(J^P)=0^-(1^+)$ channel due to significant isospin breaking by the Coulomb force. This channel, not readily accessible by other means, may contain near-threshold molecular meson-antimeson resonances --- isoscalar analogs of the isovector states $Z_b(10610)$ and $Z_b(10650)$.  

\end{abstract}
\end{titlepage}

The resonances $Z_b=Z_b(10610)$ and $Z_b'=Z_b(10650)$ found~\cite{bellez} in the transitions $\yp(5S) \to \pi^\pm \, Z_b^\mp$ are very close in mass to the respective thresholds $B \bar B^*$ and $B^* \bar B^*$ and are naturally interpreted~\cite{bgmmv} as $S$-wave molecular states in these meson-antimeson channels with quantum numbers $I^G(J^P) = 1^+ (1^+)$. In particular, the molecular interpretation, unlike alternative schemes~\cite{dos,gczc,nnr,ahw}, appears to fully agree with the observed pattern of masses and widths of the new resonances as well as with the interference pattern between the resonances in all their known decay channels~\cite{bgmmv}. The isospin of both $Z_b$ resonances is equal to one. Therefore there should also exist neutral counterparts of the observed charged states,  $Z_b^0$ and $Z_b'^0$, which are produced in the processes $\yp(5S) \to \pi^0 \, Z_b^0$. Furthermore, one can also expect~\cite{bgmmv,mvw} existence of isovector states with negative $G$ parity near the $B \bar B$, $B \bar B^*$ and $B^* \bar B^*$ thresholds as well as of isoscalar states with both positive and negative $C$ parity. The $I^G=1^-$ states and the $C$-even isoscalar resonances are in principle accessible in radiative transitions from $\Upsilon(5S)$, although it is not clear whether the existing data are sensitive enough to such processes. The $C$-odd isoscalar states however cannot be produced in radiative decays of $\yp(5S)$, neither isospin-conserving hadronic transitions from $\yp(5S)$ are kinematically possible. The latter states are direct isoscalar analogs of the isovector resonances $Z_b$ and $Z_b'$, and can be expected to exist near the same $B \bar B^*$ and $B^* \bar B^*$ thresholds. Furthermore, the isoscalar states can mix with pure bottomonium, and thus their study would possibly shed some light on both the significance of such mixing and of an isospin-dependent interaction between the heavy $B^{(*)}$ mesons. The same mixing can generally facilitate a production of these resonances at LHC~\cite{bgmmv}, however at present it appears impossible to reliably estimate the corresponding cross section and thus to assess the feasibility of their study using high-energy hadronic collisions. 

In future, a study of the $C$-odd isoscalar threshold states ($X_b$) of $S$-wave $B^{(*)}$ meson pairs might become possible if energies above approximately 11.4\,GeV could be attained with high luminosity in $e^+e^-$ collisions and if there exist suitable analogs of the $\yp(5S)$ at those higher masses that would decay to e.g. $\eta + X_b$. Given that both these possibilities are yet somewhat remote, it would be interesting if an insight into the near-threshold interaction between the $B^{(*)}$ mesons in the $I^G(J^P)=0^-(1^+)$ channel could be gained from the existing or forthcoming $e^+e^-$ data at the $\yp(5S)$ resonance. The purpose of the present paper is to argue that such an insight can be provided by the ratio $R^{c/n}$ of the yield of pairs of charged and neutral $B^{(*)}$ mesons in the decays $\yp(5S) \to \pi^0 \, B^{(*)} \bar B^{(*)}$ with very slow heavy mesons. Namely, at a small c.m. velocity $v$ of the heavy mesons the ratio $R^{c/n}$ is modified by the Coulomb interaction between the charged mesons, and the parameter for this modification is $\alpha/v$. The specific behavior of the Coulomb effect for the pairs produced by an isovector source, the decay $\yp(5S) \to \pi^0 \, B^{(*)} \bar B^{(*)}$, is determined by the strong scattering amplitude in the {\it isoscalar} channel~\cite{dlorv}. Furthermore, at near-threshold c.m. energy only $C$-odd $S$-wave states of the heavy meson pairs do contribute in the discussed decays of $\yp(5S)$ and such states are only possible at the $B \bar B^*$ and $B^* \bar B^*$ thresholds and have the quantum numbers $I^G(J^P)=0^-(1^+)$.

The molecular picture implies that the $Z_b$ resonances should have a strong coupling to the respective channels $B \bar B^*+c.c.$ and $B^* \bar B^*$ and should thus enhance the production of the meson pairs in the near threshold region in the decays $\yp(5S) \to \pi \, B^{(*)} \bar B^{(*)}$. Clearly, such an enhancement should significantly facilitate the discussed here study of the process $\yp(5S) \to \pi^0 \, B^{(*)} \bar B^{(*)}$. 

The discussed treatment of the Coulomb effects in the ratio $R^{c/n}$ is based on the standard assumptions about the strong interaction (see e.g. in the textbook \cite{ll}). Namely, it is assumed that the strong force between heavy mesons is confined to a region with size $a$, beyond which the wave functions are described by free motion which is modified for charged mesons when the Coulomb interaction is taken into account. In the absence of the Coulomb force the stationary-state wave function of the meson pair splits into independent radial functions in the two isotopic channels, $R_0(r)=\phi_0(r)/r$ and $R_1(r)=\phi_1(r)/r$, and neglecting an absorption to inelastic channels each of these functions is written in the $S$ wave in the form
\be
\left. \phi_{0,1}(r) \right |_{r > a}=b_{0,1}^* \, f_+(r) + b_{0,1} \, f_-(r)~,
\label{swf}
\ee
with $f_-(r)$ ($f_+ = f_-^*$) describing an incoming (outgoing) wave. For the free motion at momentum $p$ in the $S$-wave one has $f_\pm(r) = \exp ( \pm i p r)$. The coefficients $b_{0,1}$ are determined by the scattering phases $\delta_{0,1}$ in the corresponding isotopic channels. In the $S$-wave the relation reads as
\be
\exp \left ( 2 i \delta \right ) = - b^* / b~.
\label{dph}
\ee

The corresponding wave functions of the pairs of charged mesons, $B^{(*)+}  B^{(*)-}$: $\phi_c(r)$, and of the neutral ones, $B^{(*)0} \bar B^{(*)0}$: $\phi_n(r)$, are given by the sum and the difference of the functions for isotopic eigenstates: $\phi_c = \phi_0 + \phi_1$,  $\phi_n = \phi_0 - \phi_1$.

When the meson pairs are produced by an isotopically pure source, i.e. with $I=0$, or $I=1$, and in the zeroth order in the isospin-breaking Coulomb interaction, the relative yield of pairs of charged and neutral mesons is equal to one: $R^{c/n}_0=1$. The Coulomb interaction between the charged mesons breaks this relation, and the first order correction for an isotopically pure source depends on the strong scattering in the {\it opposite} isotopic channel~\cite{dlorv}. In the discussed  decay $\yp(5S) \to \pi^0 \, B^{(*)} \bar B^{(*)}$ the heavy mesons are produced in the $I=1$ state, and a straigtforward adaptation of the expression for the correction to $R^{c/n}$ from Ref.~\cite{dlorv} reads as
\be
R^{c/n}=1- { 1 \over v} \, {\rm Im}\left [ e^{2i \delta_0} \,
\int_a^\infty e^{2ipr}  \,
V(r) \, dr \right ]~,
\label{rcn}
\ee
where $V(r)$ is the Coulomb interaction potential between the charged heavy mesons. Clearly, in the absence of a form factor one has $V(r)=-\alpha/r$. The electromagnetic form factor of the $B^{(*)}$ meson can modify the interaction at short distances, most likely, eliminating at such distances the isotopic difference between the charged and the neutral mesons. In the present approach, for simplicity, it is assumed that the length parameter $a$ generically denotes the distance scale at which the strong interaction becomes effective and where the isotopic difference in the electromagnetic interaction becomes unimportant, and the point-like Coulomb expression for $V(r)$ is used down to $r=a$. In this case the integral takes the explicit form:
\be
R^{c/n}=1+ {\alpha \over v} {\rm Im} \left \{ e^{2i \delta_0} \, \left [ {i\, \pi \over 2} - i \, {\rm Si} (2 \, p \, a) - {\rm Ci}(2 \, p \, a) \right ] \right \}~.
\label{rcns}
\ee
One can readily see that in the absence of interaction in the $I=0$ channel, $\delta_0 \to 0$ the familiar formula $R^{c/n} = 1+ \pi \alpha /2 v$ is recovered from Eq.(\ref{rcns}) in the limit $a \to 0$.

\begin{figure}[ht]
  \begin{center}
    \leavevmode
    \epsfxsize=16cm
    \epsfbox{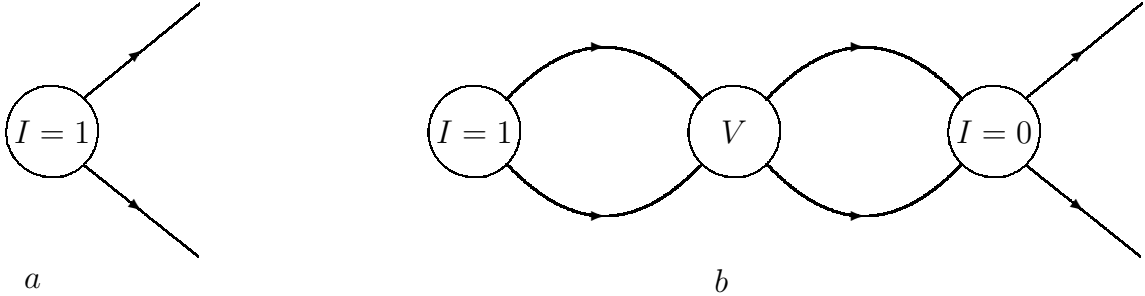}
    \caption{Graphical representation of the processes described by Eq.(\ref{rcn}). The production of free heavy mesons by an insovector source ($a$) and the first Coulomb correction with rescattering in the isotopic $I=0$ state ($b$).}
  \end{center}
\end{figure}

The detailed derivation of the expression equivalent to  Eq.(\ref{rcn}) is described in Ref.~\cite{dlorv}. It can be graphically illustrated as shown in Fig.~1. The meson pair is produced in the $I=1$ isotopic state (the leftmost circle in the zeroth order ($a$) and in the first order ($b$) in the Coulomb interaction $V$). In the zeroth order in $V$ only an outgoing wave $f_+(r)$ is present. In the first order the emitted outgoing wave in the channel with charged mesons rescatters due to the potential $V$ into an incoming wave, whose amplitude at $r=a$ is found as
\be
\delta f_c(r) \left. \right |_{r \to a}  = \xi \,
f_-(r) 
\label{dchia} 
\ee 
with 
\be \xi = -{i \over 2 v}\,
\int_a^\infty V(r') \, \left [f_+(r') \right]^2 \, dr'~.
\label{etaex}
\ee
This extra contribution to the incoming wave function of the charged mesons then should match the outgoing wave in the $I=0$ channel (the rightmost circle in Fig.~1$b$), which matching, according to Eq.(\ref{swf}), is described by the factor $\exp (2 i \delta_0)$. The interference of the resulting outgoing $I=0$ wave with the one in the zeroth order thus gives rise to Eq.(\ref{rcn}). 

Using the illustration in Fig.~1 one can in fact generalize Eq.(\ref{rcn}) for the case where there is an absorption due to inelastic processes of the heavy mesons in the $I=0$ state. The effect of such absorption can be written in terms of a real elasticity factor $0< \eta \le 1$ in diagonal element of the $I=0$ $S$ matrix: $\exp( 2 i \delta_0) \to \eta \, \exp( 2 i \delta_0)$, which results in a re-scaling by $\eta$ of the Coulomb term in Eq.(\ref{rcn}).  

One can readily notice that the described approach and the resulting relation (\ref{rcns}) can be applied only in a limited range of the energy of the heavy meson pair above the threshold. Indeed, the details of the interplay of the Coulomb and the strong interaction at short distances are not important, and can be coded in terms of one scale $a$,  only as long as the product $p a$ is small. Setting (somewhat arbitrarily) the strong interaction scale as $a \approx 1\,$fm, limits the range of the c.m. energy $E$ above the threshold to not exceeding approximately $8\,$MeV. On the other side, the limitation is set by the use of the linear approximation in the Coulomb interaction, described by the parameter $\alpha/v$. The higher order contributions are reasonably under control at approximately $E > 2\,$MeV. Both these limitations can be relaxed in a more refined treatment, that may become relevant if and when experimental data become available. 

Although the discussed range of the energy, where Eq.(\ref{rcn}) can be applied, is quite limited, the deviation of the ratio $R^{c/n}$ can display a substantial variation over this range and is also significantly sensitive to the phase $\delta_0$ as illustrated in Fig.~2. In particular, if there is an isoscalar resonance near the threshold, it is likely to be revealed by the characteristic variation of $R^{c/n}$ with energy. 
\begin{figure}[ht]
  \begin{center}
    \leavevmode
    \epsfxsize=12cm
    \epsfbox{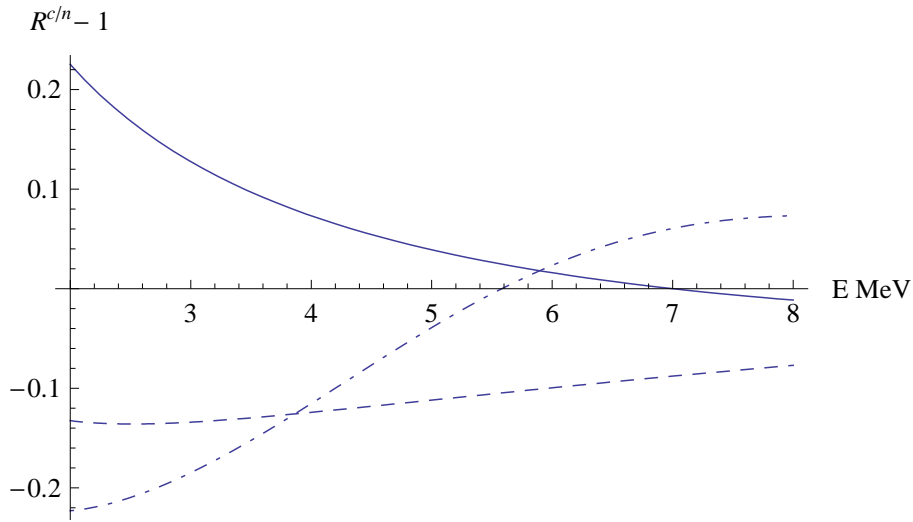}
    \caption{Sample behavior of the Coulomb effect in the ratio $R^{c/n}$ at $a=1\,$fm for different assumed forms of the $I=0$ strong scattering amplitude: no strong scattering, $\delta_0=0$ (solid), fixed scattering phase $\delta_0=45^\circ$ (dashed), and in a presence of an $I=0$ resonance at $E_0=5\,$MeV with the width $\Gamma=10\,$MeV (dotdashed).}
  \end{center}
\end{figure}

If the masses of such resonances, $X_b$ and $X_b'$, are within few MeV from respectively the threshold for $B \bar B^*$ and $B^* \bar B^*$, there in principle can exist another revealing process due to the Coulomb isospin breaking. Namely, such resonances should have decays into $\eta+\Upsilon(1S,2S)$ and $\eta + h_b(1P)$. Using the (inelastic) coupling to these final states in the $I=0$ amplitude in Fig.~1$b$, one concludes that there should be observable processes of the type $\Upsilon(5S) \to \pi^0 \, \eta + (bottomonium)$. The rate of such processes can be very approximately estimated as
\be
{\Gamma[\Upsilon(5S) \to \pi^0 \, \eta + (bottomonium)] \over \Gamma[\Upsilon(5S) \to \pi \, \pi + (bottomonium)]} \sim \left ( {\alpha \over v_0} \right )^2~  
\label{pieta}
\ee
where $v_0$ is the c.m. velocity of the heavy mesons at the resonance energy. The parameter in the r.h.s. of this estimate may turn out to be not small if the resonance is sufficiently close to the threshold.

It can be also mentioned that another isospin-breaking effect due to the Coulomb interaction between the heavy mesons is a mass difference between the known charged $Z_b^\pm$, $Z_b'^\pm$ resonances and their neutral isospin partners, $Z_b^0$ and $Z_b'^0$. However due to a large spatial size of these molecular states the isotopic mass shift is likely less than approximately 1\,MeV.

 This work is supported, in part, by the DOE grant DE-FG02-94ER40823.

\end{document}